# Frequency Scaling Laws for Flat Plate Wing Active Separation Control

Stefan Vey

*Rolls-Royce Deutschland, 15827 Blankenfelde-Mahlow, Germany*

Christian Oliver Paschereit

*Technische Universität Berlin, Müller-Breslau-Str. 8, 10623 Berlin, Germany*

David Greenblatt

*Technion – Israel Institute of Technology, Technion City, Haifa 3200003, Israel*

## Abstract

Dimensionless frequency scaling laws for active separation control on flat-plate wings, using dielectric barrier discharge plasma actuators, were examined on the basis of maximum increases to lift coefficient, and compared with hovering insect wing-flapping frequencies. Data for a range of angles of attack (24° to 32°), Reynolds numbers $(3 \times 10^3$ to $20 \times 10^3)$, and semispan wing aspect ratios (0.75 to ∞), collapsed best when scaled with the streamwise-directed height of the chord length. The "forcing Strouhal number" that produced the largest lift coefficient increments, equal to 26±0.04, was linked to the conventional bluff-body Strouhal number by recognizing that drag and lift on flat plate wings are directly proportional at a fixed angle of attack. Flowfield measurements, to determine the time-averaged separation bubble height and local velocity at separation, showed that the universal Strouhal number of approximately 0.16±0.01—developed for bluff-body and separation bubble vortex shedding—can be further generalized to active separation control, when the natural shedding frequency is substituted by the forcing frequency. Insect wing flapping frequencies in hover were examined on the basis of Strouhal number scaling and corresponded reasonably well to the optimum forcing Strouhal number range, although angle of attack estimates were a source of uncertainty. Furthermore, universal Strouhal number scaling can only be validated with accurate separation bubble height and separation velocity measurements.

# Nomenclature

| | | |
|---|---|---|
| $A$ | = | wing flapping amplitude, m |
| $C_D$ | = | wing drag coefficient $F_D / \tfrac{1}{2} \rho V_{ref}^2 S$ |
| $C_d$ | = | airfoil drag coefficient $F_d / \tfrac{1}{2} \rho V_{ref}^2 c$ |
| $C_L$ | = | wing lift coefficient $F_D / \tfrac{1}{2} \rho V_{ref}^2 S$ |
| $C_l$ | = | airfoil drag coefficient $F_l / \tfrac{1}{2} \rho V_{ref}^2 c$ |
| $C_{p,sep}$ | = | pressure coefficient at separation |
| $C_\mu$ | = | plasma force coefficient, $F_p / \tfrac{1}{2} \rho V_{ref}^2 c$ |
| $c$ | = | airfoil chord-length, m |
| $\bar{c}$ | = | wing average chord-length, m |
| $D$ | = | projected bluff-body height, m |
| $D'$ | = | distance between free-streamlines, m |
| $F_p$ | = | DBD plasma body force, N/m |
| $F^+$ | = | forcing reduced frequency |
| $f_f$ | = | wing flapping frequency, Hz |
| $f_p$ | = | perturbation frequency, Hz |
| $f_{sh}$ | = | vortex shedding frequency, Hz |
| $h$ | = | transverse height, $D/2$, m |
| $h_b$ | = | bubble height, $D'/2$, m |
| $k$ | = | reduced frequency, $\pi f_f \bar{c} / \bar{V}$ |
| $R$ | = | wing semispan length, m |
| $S$ | = | wing area, m$^2$ |
| $Re$ | = | Reynolds number, $U_\infty c / \nu$ or $\bar{V} \bar{c} / \nu$ |
| $St$ | = | Strouhal number, $f_{sh} D / U_\infty$ |
| $St^*$ | = | biomechanics Strouhal number, $f_f A / V_{ref}$ |
| $St_U$ | = | universal Strouhal number, $f_{sh} D' / U_{sep}$ |
| $St^+$ | = | forcing Strouhal number, $f_p D / U_\infty$ |
| $St_U^+$ | = | universal forcing Strouhal number, $f_p D' / U_{sep}$ |
| $U_{sep}$ | = | local velocity at separation, m/s |
| $U_\infty$ | = | freestream wind speed, m/s |
| $\bar{V}$ | = | average wingtip speed, $2 f_f \Phi R$, m/s |
| $V_{ref}$ | = | reference velocity, $U_\infty$ or $\bar{V}$, m/s |
| $x, y, z$ | = | cartesian coordinates, m |
| $x_a$ | = | linear oscillations amplitude, m |
| $x_b$ | = | separation bubble length, m |
| $\alpha$ | = | angle of attack, ° |
| $\alpha_s$ | = | static stall angle of attack, ° |
| $\Lambda$ | = | aspect ratio, $2R/\bar{c}$ |
| $\Lambda_s$ | = | semispan aspect ratio, $R/\bar{c}$ |
| $\Phi$ | = | flapping stroke amplitude, rad |

# 1   Introduction

The development of transitory high-lift on flat-plate airfoils and wings can be achieved by exploiting unsteady aerodynamics mechanisms. For example, rapidly pitching airfoils can produce up to 3.5 times greater lift coefficients than their steady counterparts via the generation of large leading-edge vortices (LEVs) [1-4]. Similar LEVs are generated periodically by flapping-wing flyers—like insects, birds and bats—for hovering, maneuvering, and forward flight [5-8], and are the primary unsteady aerodynamic mechanism responsible for lift generation. Hence the engineering problem of efficient lift generation for flapping wing micro aerial vehicles (FWMAVs), can be viewed as an exercise in active LEV control [9,10]. FWMAVs that operate at Reynolds numbers below about $20 \times 10^3$ [8,11-16] mimic the platelike geometry, kinematics and flexibility of insect wings in an effort to efficiently generate and exploit LEVs. A central challenge is to identify an optimum frequency range that produces the greatest lift force, for the lowest energy input. The problem can be significantly simplified if we confine our attention to the problem of hovering flight, and we assume that the wings are thin, flat and rigid.

Significant lift augmentation can also be developed on stationary, stalled, flat-plate airfoils by the artificial imposition of periodic leading-edge perturbations, which lead to a train of LEVs [17]. When the perturbation amplitudes are on the order of the freestream velocity—produced, for example by dielectric barrier discharge (DBD) plasma actuators—they periodically sever the leading-edge shear layer, leading to the rollup and advection of LEVs [17]. In these active separation control problems, the frequency parameter is usually characterized by a ratio of the convective and perturbation timescales [18]:

$$F^+ \triangleq \frac{f_p c}{U_\infty} \quad (1)$$

where $f_p$ is the perturbation frequency, $c$ is the airfoil or wing chord-length, and $U_\infty$ is the freestream velocity. On flat and cambered plate airfoils, optimum reduced frequencies—defined as those that produce the largest post-stall lift coefficients—typically fall in the range $0.2 \leq F^+ \leq 0.4$ [17-20]. This reduced frequency range ensures that two vortices are present above the airfoil surface at phase of the forcing cycle, while in a time-averaged sense, a large separation bubble is enclosed on the surface [17].

Insect flapping lift generation and active separation control bear a resemblance to bluff body vortex shedding [7,21]. Bluff-body shedding is commonly characterized by the ratio of convective to vortex shedding timescales, termed the Strouhal number:

$$St \triangleq \frac{f_{sh} D}{U_\infty} \qquad (2)$$

where $f_{sh}$ is the vortex shedding frequency, and $D$ is the projected height of the body in the freestream direction. Vortices are shed alternately in the wake of bluff bodies, giving rise to the well-known von Kármán vortex street. The best-known example is that of two-dimensional cylinders, where the vortex shedding frequencies collapse approximately to a single value $St \approx 0.21$, in the subcritical Reynolds number range of ~500 to ~$10^5$ [22]. Changes to the bluff body's geometric shape, e.g., circular cylinder versus normal flat plate, but with the same projected height, produce a change in the vortex shedding frequency. Roshko [23] showed, however, that a generalized or "universal" Strouhal number can be developed to characterize vortex shedding by a single dimensionless number, irrespective of the body's geometry.

This research has two main objectives. The first is to determine scaling laws for active separation control on flat plate wings and demonstrate their consistency with the vortex-shedding Strouhal number definition. The second is to compare this scaling with that associated with insect flapping-wing hovering flight. To attain these objectives, we first review the assumptions associated with the universal Strouhal number and then explain the relationship between bluff-body shedding and flat plate wing lift (section 2). Next, we describe the semispan and finite wing experiments (section 3), which augment our existing airfoil data sets presented in [17-20]. Non-dimensional data sets are then discussed to illustrate the main scaling laws (section 4.1) and, in particular, data consistency with the universal Strouhal number (section 4.2). This is followed by a comparison to dimensionless Strouhal number ranges associated with insect wing flapping in hovering flight (section 4.3), followed by the main conclusions (section 5).

## 2 The Universal Strouhal Number

### 2.1 Bluff Bodies & Separation Bubbles

The concept of a universal Strouhal number for arbitrary bluff bodies, first proposed by Roshko [23], uses the distance between the free streamlines (of potential flow) $D'$ and the velocity magnitude at separation $U_{sep}$ as the scaling parameters. These quantities are shown schematically for a circular cylinder and normal flat plate in the top row of Figure 1. The pressure distribution, and hence the drag on bluff bodies can then be computed by specifying only the base pressure coefficient $C_{p,sep} = 1 - (U_{sep}/U_\infty)^2$, which facilitates the calculation of $D'/D$ by using a modification to Kirchhoff's method. The relationship between the traditional and the universal Strouhal numbers is thus:

$$St_U \triangleq \frac{f_{sh}D'}{U_s} = St \frac{U_\infty}{U_{sep}} \frac{D'}{D} \tag{3}$$

where $St_U$ lies between 0.15 and 0.17, with a median value of approximately 0.16 for a wide variety of different two-dimensional bluff-bodies. It is interesting to note that a similar result can be obtained based on a simple energy balance [24], namely, if we equate the mechanical energy of the fluid layer along a streamline $\tfrac{1}{2}(2\pi f_{sh} D')^2$ with the available kinetic energy $\tfrac{1}{2} U_{sep}^2$, we get:

$$\left( \frac{f_{sh} D'}{U_{sep}} \right)_{energy} = \frac{1}{2\pi} \approx 0.16 \tag{4}$$

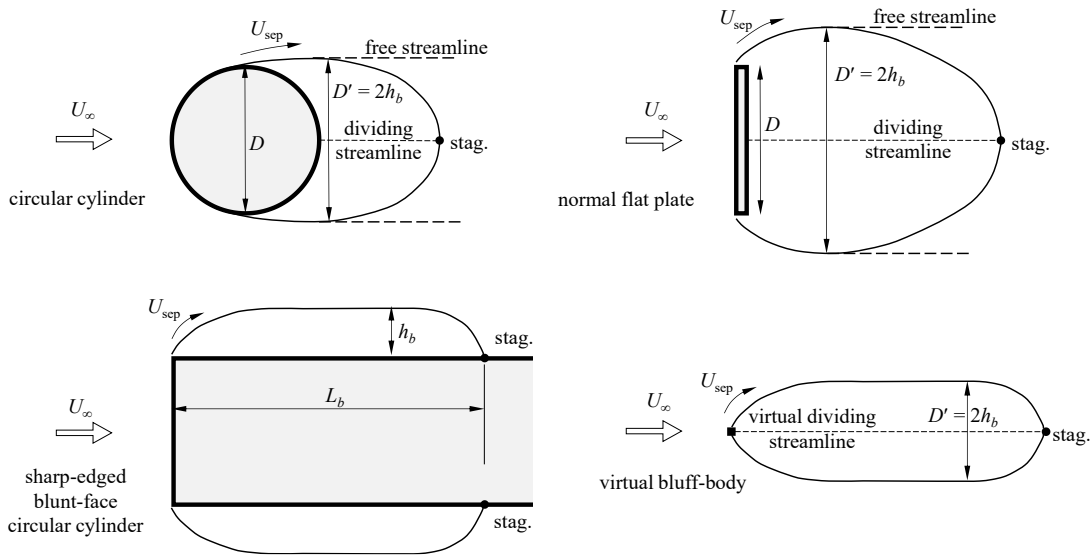

Figure 1. Mean streamline shapes and parameters used in Roshko's universal Strouhal number $f_{sh}D'/U_{sep} \approx 0.16$ [23,26,27].

Strouhal number universality can be extended to flows with solid surfaces by recognizing that both a stagnation point and a line of symmetry exist, in a mean sense, aft of a bluff-body. This is shown schematically in the top row of Figure 1, and should be contrasted with Mabey's [25] empirical correlation $f_{sh} x_b / U_\infty = 0.65 \pm 0.15$. If the two sides of the wake are images of one another, without any loss of generality we can state that the wake width and bubble height are related by $h_b = D'/2$. If the line of symmetry, or mean streamline, is replaced by a solid wall, then the same universal scaling should apply. Sigurdson and Roshko [26,27] showed this to be true on an axially mounted cylinder, where $f_{sh} h_b / U_{sep} \approx 0.08$ (i.e., $St_U \approx 0.16$), shown in Figure 1 (bottom left). Here the physical problem is modelled as the wake behind an infinitely small point (virtual bluff body), as shown in Figure 1 (bottom right), where the solid surface is represented by a virtual mean streamline. This result is consistent with a variety of separation bubbles, where it can be seen that $f_{sh} h_b / U_{sep} = 0.075 \pm 0.015$ [27], and is related to Mabey's [25] correlation by $(h_b / x_b)(U_{sep} / U_\infty) \approx 0.115$. A drawback of this scaling is that $h_b$ and $U_{sep}$ are often not conveniently measurable quantities.

## 2.2 Bluff-Body Drag and Flat-Plate Wing Lift

In this section, we address the relationship between drag and vortex shedding, for reasons that will become clear below. While it cannot be proven from first principles that the state of vortex shedding results in the maximum drag on a particular bluff body at a given Reynolds number, shedding increases the rate of kinetic energy dissipation, or entropy production, in the wake—and thereby increases drag. When shedding is weakened or its effects are reduced, for example, by inserting splitter plates along the dividing streamline (see Figure 1) the mean drag decreases [28]. The attainment of maximum, or at least high, drag by vortex shedding is a critically important link between bluff body shedding and flat-plate active separation control. This is because increases to post-stall flat-plate drag, due to the active generation of LEVs, are linearly proportional to changes in lift at a given angle of attack, based on the following simple reasoning: At Reynolds numbers greater than a few hundred, the normal (pressure) forces on a plate $F_N$ vastly exceed the tangential (viscous) forces $F_T$; hence $F_L = F_N \cos\alpha$, $F_D = F_N \sin\alpha$, and thus $F_D = F_L \tan\alpha$. These arguments are equally valid for flat plate airfoils and finite wings, but they are not valid for conventional thin or thick airfoils and wings because, in general, lift and drag are not linearly, or even proportionally, related. Flat-plate wings have the additional advantage that they have fixed (leading-edge and tip) separation points, which renders their aerodynamic loading virtually Reynolds number independent in the practically important $10^2$ to $10^5$ Reynolds number range.

# 3　　Experimental Configurations

Results presented in the following sections were obtained on both semispan and finite wings, in the range $18 \times 10^3 \leq Re \leq 50 \times 10^3$. Semispan wing experiments were performed in a closed-return wind tunnel with a test section of 0.28 m × 0.4 m (height × width), with $U_{\infty,max} = 25$ m/s and a turbulence level $Tu = 0.15\%$. Three semispan wings ($c = 150$ mm, $t/c = 1.33\%$) of $\Lambda_s \triangleq R/c = 0.75$, 1.00, and 1.27 were manufactured from glass-reinforced epoxy laminate with rounded edges. Each wing was mounted vertically, 0.35 hydraulic diameters downstream of the nozzle exit, on a custom two-component load balance as shown in Figure 2 (left), with a 0.5 mm gap between the wing root and wall. Flow speed and fluid properties were based on a PT 100 temperature sensors and a Pitot-static probe mounted near the opposite wall upstream of the wings.

Finite wing experiments were performed in a closed-return wind tunnel with a test section of 1.4 m × 2.0 m, $U_{\infty,max} = 60$ m/s and a maximum turbulence level $Tu = 0.5\%$. A finite wing ($c = 150$ mm, $\Lambda = 2.66$, $t/c = 2\%$) was attached to a mounting sting, that incorporated a normal force sensor, and was installed vertically in the test section (Figure 2, right). This assembly was attached to a vertical support beam which was rigidly attached to a turn table. Two-dimensional particle image velocimetry (PIV) measurements were made using a 160 mJ Nd:YAG double-pulsed laser with a wavelength of 532 nm. The laser beam was guided to the test section by a light-arm which also incorporated light-sheet optics, which were aligned with the wing centerline. Two CCD-cameras (2048 × 2048 pixel resolution), with accompanying Scheimpflug adapters were employed to acquire 942 image-pairs for each configuration. Reflections were reduced by covering the wing the surface with a fluorescent foil and then bandpass-filtering the images outside of the 532 nm range. Phase-averaged and long-time average measurements were made.

All wing leading-edges were equipped with flush-mounted DBD plasma actuators, of identical design, along their entire span. The exposed and covered electrodes were cut from 70 µm thick tinned copper tape, with widths of 5 mm and 10 mm respectively, and separated by three layers of 50-µm-thick Kapton® tape (see inset in Figure 2, left). The trailing and leading edges of the exposed and covered electrodes were set at the wings' leading-edges. A Miniplus 1® high-voltage generator was employed at ionization frequencies of between 4.0 kHz to 4.75 kHz, and peak-to-peak voltages between 8 kV to 12 kV. Direct body force calibrations were performed by measuring the actuator reaction force using a Vibra AJ-220E laboratory scale, described in [19]. For aerodynamics experiments, the actuators were pulse modulated at frequencies $f_p < 100$ Hz, with duty cycles between 5% and 20%. Addition details regarding the data experimental setups and data acquisition can be found in [29-32].

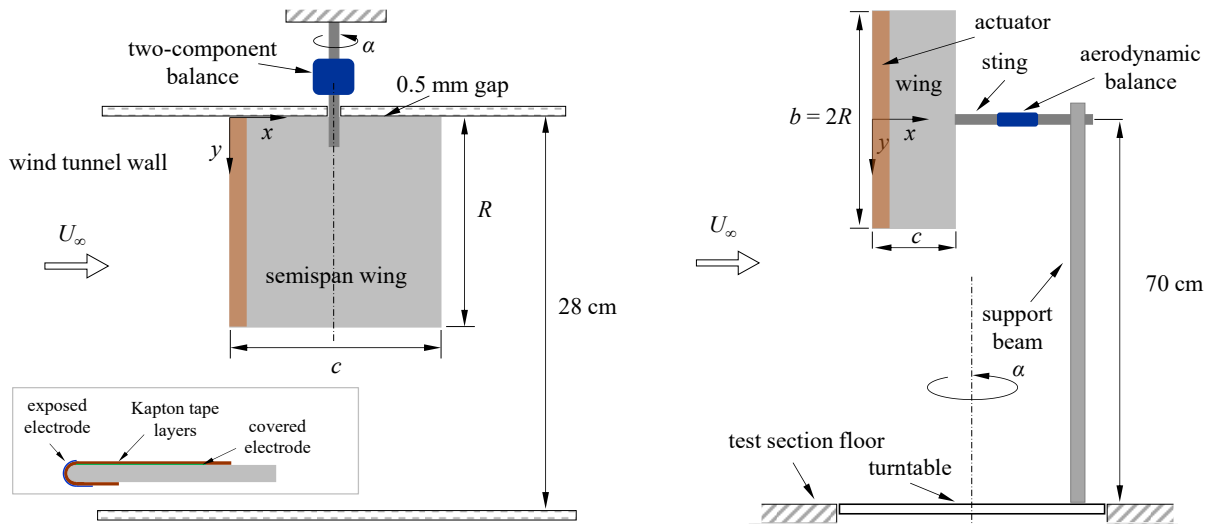

Figure 2. Schematics of the experimental setups for: the semispan wings with $\Lambda_s = 0.75, 1.00,$ and 1.27 (left); and the finite wing with $\Lambda = 2.66$. Left inset: leading-edge detail of the DBD plasma actuators.

## 4   Experimental Results and Scaling

Initial baseline (no perturbations) and active separation control experiments were performed on the three semispan wings and the lift coefficient ($C_L = F_L / \tfrac{1}{2} \rho U_\infty^2 Rc$) versus angle-of-attack results are shown in Figure 3 at $Re = 20 \times 10^3$. The control parameters selected, namely $F^+ = 0.24$ and $C_\mu = 0.025$, were based on our prior flat plate airfoil results described in [17,20]. As anticipated, the wings with small aspect ratios produced smaller $dC_L / d\alpha$ slopes due to the stronger induced downwash from the wingtip vortices. Also consistent with expectations, baseline $C_{L,\max}$ increased and $\alpha_s$ decreased from $\alpha = 16°$ and $21°$ with increasing aspect ratio, again due to the dominant role played by the wingtip vortices.

The introduction of large-amplitude leading-edge perturbations had a pronounced effect on both $C_{L,\max}$ (increases of between 56% and 78%) and $\Delta\alpha_s$ (increases of between 14° and 16°). Note that here we cite the total momentum coefficient $C_\mu$ and not the duty-cycle-dependent quantity $\langle C_\mu \rangle \triangleq C_\mu(\text{d.c.})$ as we have in previous studies [17-20]. The reason for this is that we observed no measurable differences in lift coefficient measurements for duty cycles between 5% and 20%; duty cycle changes only result in a proportional change in power input. The results here represent large amplitude perturbations, i.e., perturbation velocities are on the order of the freestream velocity. The semispan lift coefficient increases were somewhat larger than our airfoil results, namely $\Delta C_{l,\max} \leq 50\%$ and $\Delta\alpha_s \leq 12°$ [19] and suggest that a positive interaction exits between the periodically generated LEVs and the tip vortex. These results are consistent with other finite wing studies employing a conventional (NACA 0015) profiles [33,21].

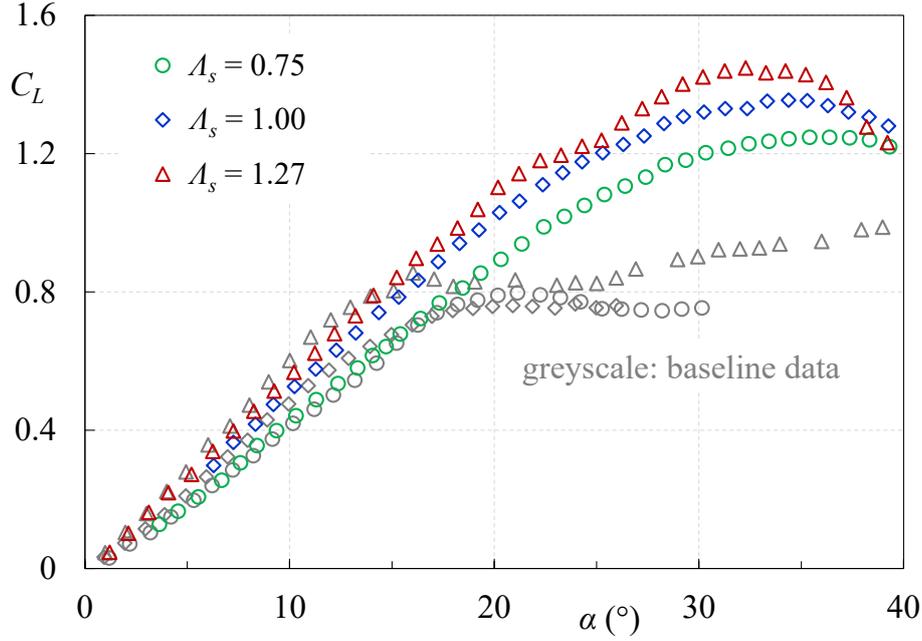

Figure 3. Baseline and active separation control results controlled at $Re = 20 \times 10^3$ under the conditions $F^+ = 0.24$ and $C_\mu = 0.025$.

## 4.1 Conventional Strouhal Number Scaling

To determine the optimum frequency ranges, each wing was placed at $\alpha = 32°$ and its frequency response—i.e., changes in lift coefficient as a function of reduced frequency—was determined. Lift coefficient results for the three semispan wings are shown in Figure 4 (left). These results are broadly similar to airfoil results [17,20] in that the changes in lift coefficient are significant and that the peak values are observed in the approximate range $0.2 \leq F^+ \leq 0.4$. The main differences between the wings are that the lowest aspect ratio wing ($\Lambda_s = 0.75$) produces a smaller increase in peak lift coefficient, while the largest aspect ratio wing ($\Lambda_s = 1.27$) exhibits a more pronounced peak. To determine if this peak is angle-of-attack independent, additional experiments were performed at $\alpha = 26°$, $29°$, and $32°$ (see Figure 4, right). Similar frequency dependence is evident at all angles, but the peak lift coefficient values occur at successively lower reduced frequencies with increasing angle of attack.

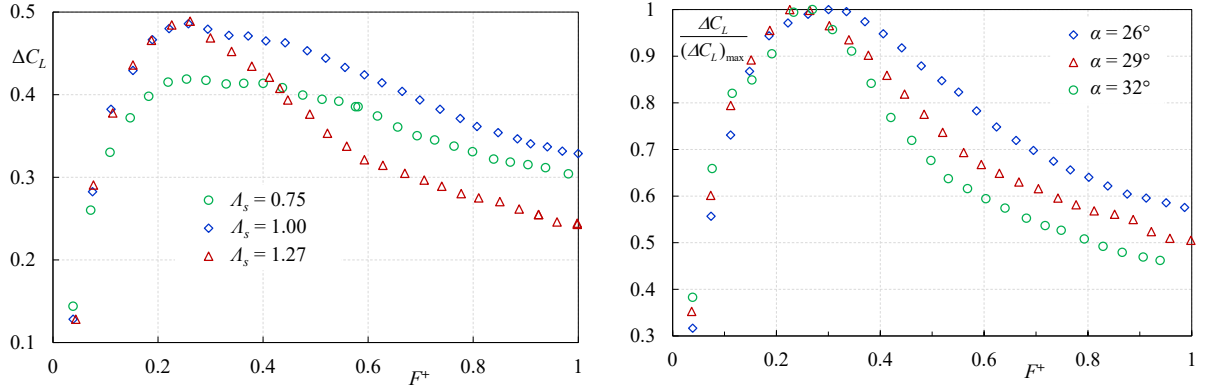

Figure 4. Changes to post-stall lift coefficient for different aspect ratio wings at $\alpha = 29°$ (left) and normalized changes for the $\Lambda_s = 1.27$ wing at different angles of attack as a function of reduced frequency.

Even though the data collapse reasonably well on the basis of $F^+$ (Figure 4), the correspondence can be improved by non-dimensionalizing the data with respect to the projected height, consistent with the Strouhal number definition. A more extensive comparison between $F^+$ and the forcing Strouhal number, is defined in general as:

$$St^+ \triangleq \frac{f_p D}{U_\infty} = \frac{f_p[2(c\sin\alpha + h_G \cos\alpha)]}{U_\infty} \qquad (5)$$

shown in Figure 5. (The "+" superscript distinguishes this Strouhal number with that associated with natural vortex shedding.) This comparison encompasses the semispan wings described above, together with airfoil data with different perturbation amplitudes, including airfoils with large Gurney flaps ($h_G/c = 10\%$ and 20% [20]), for the range $3,000 \leq Re \leq 20,000$. The forcing Strouhal number scaling produces a sharper peak than that of the reduced frequency, namely $F^+ = 0.28 \pm 0.07$ versus $St^+ = 0.26 \pm 0.04$. There is also an improved collapse of the data throughout the frequency range, apart from airfoil data with relatively small perturbation amplitudes or Gurney flaps in the low frequency range, and the two lowest aspect ratio wing data at high frequencies. It is of particular interest that the optimum reduced frequency range is barely affected within the aspect ratio range $\Lambda_R = 1.27$ to $\infty$ (although the absolute or percentage changes $C_L$ are different, as shown in Figure 4, top). This means that, although the LEVs interact with the tip vortices, the basic mechanism of lift enhancement is not affected.

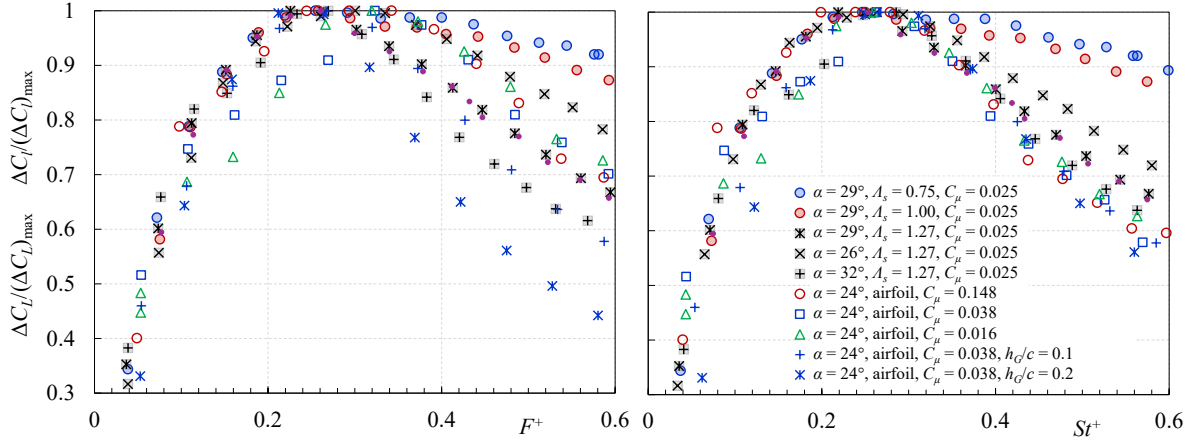

Figure 5. Normalized changes to post-stall lift coefficient for three semispan wings and airfoils with and without Gurney flaps, based on reduced frequency (left) and forcing Strouhal number (right) scaling $(3,000 \leq Re \leq 20,000)$.

## 4.2 Universal Strouhal Number Scaling

While the parameter $St^+$ is both useful and convenient, it does not provide a link to the universal Strouhal number, because neither $h_b$ nor $U_{sep}$ are known. To estimate these quantities, consider the PIV measurements made at the center-span of the $\Lambda = 2.66$ wing described in section 3. Time-averaged baseline and controlled results are shown in Figure 6, where the flowfield is visualized using line integral convolution described in [34]. The shear layers downstream of the leading- and trailing-edges are highlighted, in burgundy (negative) and azure (positive). In the baseline case, the wake is asymmetric with the mean stagnation point well aft of the airfoil and no dominant vortex shedding was detected. In the controlled case, the separation bubble along the centerline of the wing is easily discernable, as its non-dimensional length $x_b/c = 0.83$. By following the dividing streamline, it can readily be seen that the bubble height $h_b \approx x_b \sin\alpha$.

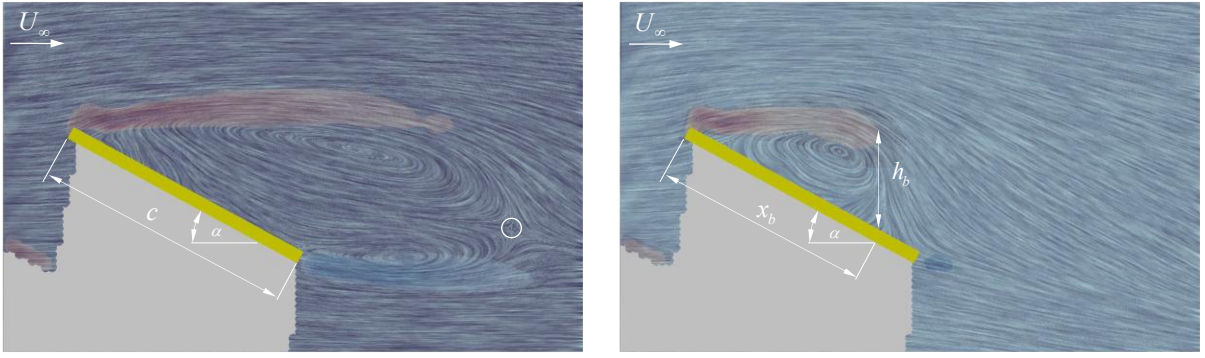

Figure 6. Time-averaged flowfield measurements at the center-span of the $\Lambda = 2.66$ wing, visualized using line integral convolution described in [34] for baseline (left) and $St^+ = 0.26$ cases $(Re = 18 \times 10^3,\ \alpha = 28°)$.

The same controlled velocity field is rendered in terms of the streamwise velocity in Figure 7, and the peak velocity ratio at the leading-edge is identified as $U_{sep}/U_\infty \approx 1.44$. Combining the bubble height estimation obtained above, together with the velocity ratio, yields:

$$St_U^+ \triangleq \frac{f_p(2h_b)}{U_{sep}} = St^+ \left(\frac{x_b}{c}\right)\left(\frac{U_\infty}{U_{sep}}\right) \approx 0.15 \qquad (6)$$

which is within Roshko's universal Strouhal number natural shedding range of $St_U = 0.16 \pm 0.01$. It is emphasized that the critical factor linking bluff-body drag and flat wing lift is the fact that drag acts as a surrogate for lift. In this sense, we are seeking to increase drag, with a byproduct being increased lift. As alluded to in section 2.2, this assumption cannot be extended to general airfoil separation control, because lift and drag are not linearly related due to non-negligible viscous effects. In fact, in most cases of active leading-edge separation control on airfoils produces a simultaneous increase in lift and reduction in drag [18]. For general airfoil and wing bubbles, the universality expressed in equation (6) only means that bubble height scales with $U_{sep}/f_p$, but tells us nothing about the lift enhancement or drag reduction, which depend on other parameters like the wing geometry and Reynolds number. For example, on a $\Lambda = 1$ wing with a NACA 0015 profile, pulsed plasma perturbations at $St^+ = 44$ ($F^+ = 50$) produced a 25% greater maximum lift coefficient increase than $St^+ = 0.75$ ($F^+ = 1$) [21]. The ratio of the velocities at separation cannot vary by more than a factor of about 2 and therefore the bubble height according to equation (6), and length, must reduce by at least an order of magnitude. These arguments are consistent with surface flow visualization [21] and computations [35]. Even though the bubble shortens, it does not scale with Mabey's parameter [25], because $f_p x_b / U_\infty \approx 0.2$ (from Figure 6, right), an hence $(h_b / x_b)(U_{sep} / U_\infty) \neq$ const.

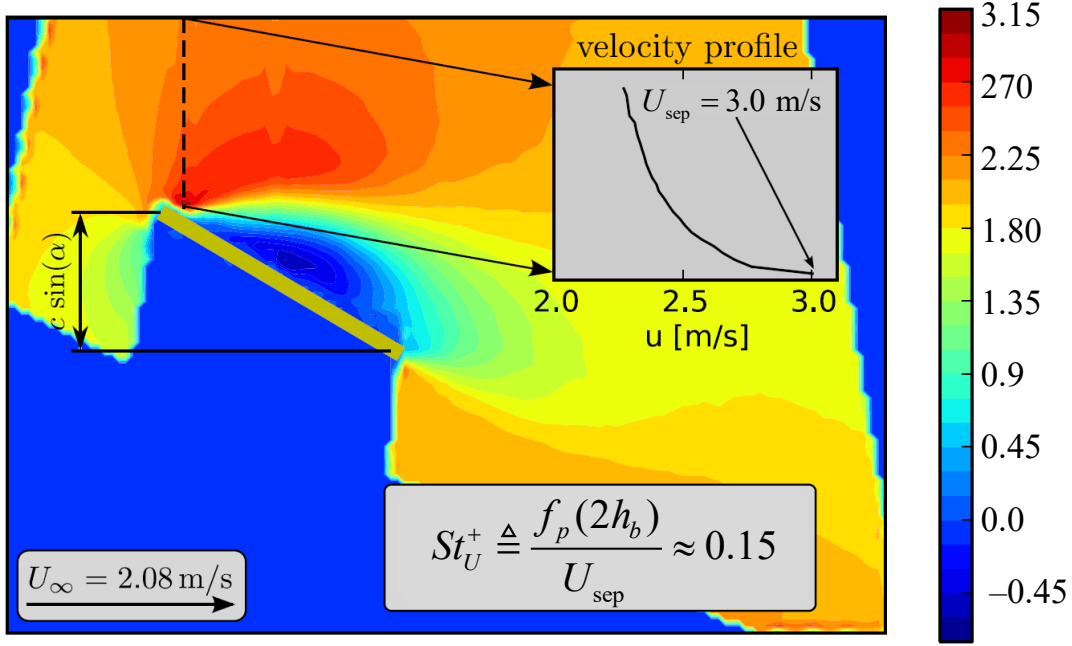

Figure 7. Time-averaged streamwise velocity component at the $\Lambda = 2.66$ wing center span, at $St^+ = 0.26$, corresponding to $St_U^+ \approx 0.15$ ($Re = 18 \times 10^3$, $\alpha = 28°$).

### 4.3 Flapping Wing Hovering Flight

Two factors prompted us to compare the scaling obtained in section 4.1 with that of insect hovering flight. First, LEVs are responsible for lift enhancement, just like they are responsible for hovering flight; and second, the optimum Strouhal number range is insensitive to aspect ratio from $\Lambda_R = 1.27$ to $\infty$, and this easily encapsulates the insect wing range of about 3 to 12. Hovering involves complex kinematic motion, often asymmetric, with deformations of the wings [6,36]. Nevertheless, the majority of the lift is generated during the translational motion due to alternating generation of LEVs and their interaction with the wingtip vortex (TV) [7,37]. To conduct a meaningful comparison to our separation control optimum Strouhal number ranges, we consider cases where $Re \geq 500$ [24], and implicitly assume the simplification proposed by [6], namely, that the semispan wing is rigid and has a constant chord-length—and therefore geometrically similar to our experimental geometries (Figure 8). We thus ignore geometric variations in planform geometry curvature, twist, etc. We further assume that the wing undergoes linear peak-to-peak oscillations at an amplitude $2x_a$, and that the rapid rotations at the end of each stroke do not contribute to lift generation (Figure 8, left) [7]. In so doing, we ignore the variation of wing velocity along the span as well as the resultant spanwise pressure gradient that arises due to insect wing flapping. We consider the wings to be linearly oscillating mechanical "actuators" that generate and manipulate the LEVs [38]—comparable to the pulsed-plasma perturbations that generated the LEVs discussed in section 4.1.

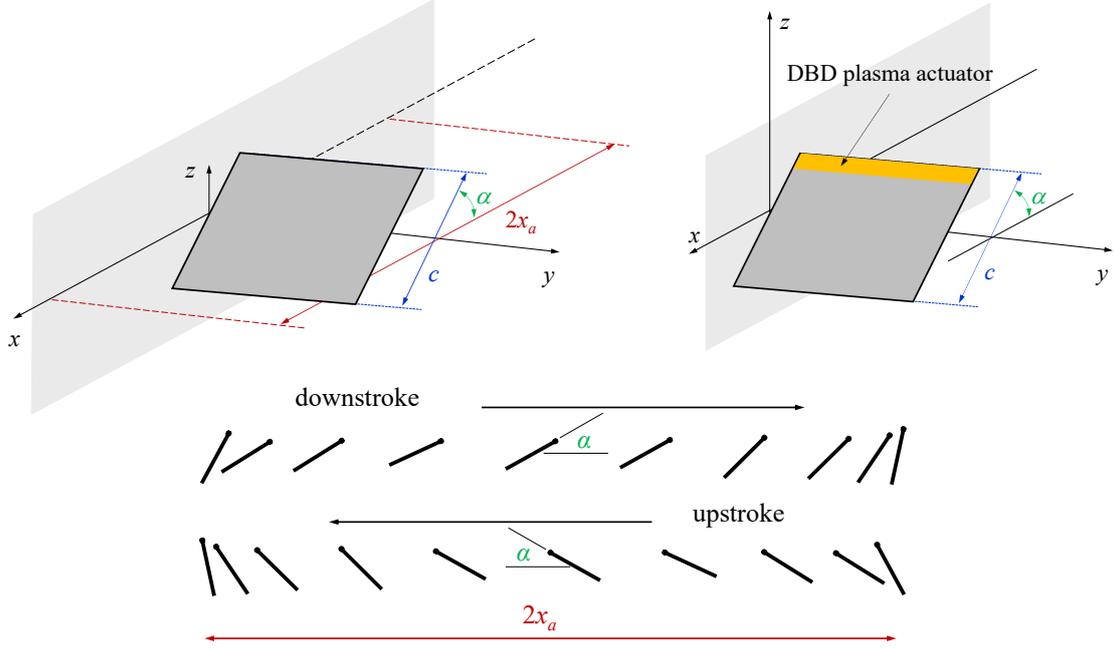

Figure 8. Simplifying assumptions for linear wing flapping in hovering flight (left and bottom) compared to the semispan experimental setup (right).

To make Strouhal number estimates relating to insect flight, it is common practice to define the average chord-length $\bar{c}$ and the peak-to-peak flapping amplitude $A = R\Phi$. The biomechanics Strouhal number is defined differently to the aerodynamics version, namely $St^* \triangleq f_f A / V_{ref}$, and forward flight (ff) and hovering flight (hf) are considered separately [6]. This definition led to the well-known observation that $0.2 \leq St^*_{ff} \leq 0.4$ for insects in forward flight [39]. However, this parameter is generally not used for hovering studies because the numerator and denominator are directly proportional. It is therefore more common to use the reduced frequency [6,40,41]:

$$k \triangleq \frac{\pi f_f \bar{c}}{V_{ref}} \quad (7)$$

where the reference velocity is usually taken as the absolute mean value at the wingtip, based on simple harmonic motion: $V_{ref} = \bar{V} = 2 f_f \Phi R$. Substitution into equation (7), yields $k_{hf} = \pi / (2\Phi\Lambda_s)$, which is independent of the flapping frequency [6]. Based on a wide variety of geometric and kinematic insect flapping wing data for $Re \geq 500$ [37] (see Appendix), it can be seen that for the vast majority of insects $k_{hf} = 0.28 \pm 0.08$, and this facilitates a direct comparison with the conventional reduced frequency for active separation control, namely:

$$F_{hf}^+ = \frac{(2f_f)\bar{c}}{V_{ref}} = \frac{2k}{\pi} = 0.18 \pm 0.05 \tag{8}$$

Note that we have rendered the flapping frequency as $2f_f$, because two flapping motions—downstroke and upstroke—occur in one cycle. This value is comparable to, but somewhat less than, the separation control optimum frequency range $F^+ = 0.28 \pm 0.07$ (see section 4.1). One possible source of discrepancy is that in insect flight, the wing root is nominally fixed, and its amplitude and speed vary linearly from root to tip. Our comparison implicitly assumes that the speed is uniform along the span, thereby overestimating $V_{ref}$ and underestimating $F_{hf}^+$. Notwithstanding these potential discrepancies, it is more likely that these reduced frequencies do not correspond because the vortex shedding arguments do not apply.

To examine data on the basis of the forcing Strouhal number, of the form defined in equation (5) of section 4.1, we write:

$$St_{hf}^+ = \frac{(2f_f)(2\bar{c}\sin\alpha)}{\bar{V}} = \frac{4k_{hf}\sin\alpha}{\pi} = \frac{2\sin\alpha}{\Phi\Lambda_s} \tag{9}$$

In the majority of insect flapping studies, the reported angles of attack are typically within ±10° [42-50] (see Table 1). Taking the minimum and maximum values of the parameters shown in Table 1, we obtain averages for the dimensionless parameter's minima and maxima, namely: $F_{hf}^+ = [0.16; 0.21]$ and $St_{hf}^+ = [0.22; 0.23]$. It is clear that the flapping wing Strouhal number scaling has a closer correspondence with the optimum separation control Strouhal number $St^+ = 0.26 \pm 0.04$ (Figure 5, right) than the reduced frequency $F^+ = 0.28 \pm 0.07$ (Figure 5, left). It should be noted, however, that the angle of attack ranges shown here are approximate, because angle-of-attack can vary within a particular stroke [51] or between upstroke and downstroke [49], and the wing can undergo twist that varies the angle along the span [6]. Furthermore, commonly cited angles of attack may be low in an average sense due to rotation of the wings, through approximately 90°, at the stroke extremities.

Notwithstanding angle of attack uncertainties, the Strouhal number scaling, which is developed here based on physical arguments, is more reliable than $F_{hf}^+$, which is just a surrogate for the commonly used $k$. One practical FWMAV consequence is that for a given aspect ratio, a unique relationship exists between stroke amplitude and angle of attack. Unfortunately, a universal Strouhal number cannot be estimated with any confidence, because direct measurements of $U_{sep}$ during flapping have not been performed. On the other hand, the bubble height $h_b$ can be estimated to first order as $\bar{c}\sin\alpha$. Dedicated and carefully controlled experiments are required to determine if universal Strouhal number scaling can be extended to flapping wing hover.

Table 1. Range of aspect ratio, flapping amplitude and nominal angle of attack ranges, together with dimensionless groups for several insect species where $Re \geq 400$.

| Species | $\Lambda_s$ | $Re$ | $\Phi$ (rad) | $\alpha$ (°) | $F_{hf}^+$ | $St_{hf}^+$ | Ref. |
|---|---|---|---|---|---|---|---|
| Eristalis tenax (hoverfly/dronefly) | 2.5 - 3.2 | ~800 - 1500 | 1.8 - 2.2 | 30 - 45 | 0.14 - 0.22 | 0.20 - 0.22 | [42] |
| Manduca sexta (hawkmoth) | 3 | 3150 | 2.2 | 30 - 40 | 0.15 | 0.15 - 0.19 | [43] |
| Calliphora (blowfly) | 3 - 3.6 | ~1300 | 2 | 40 | 0.14 - 0.17 | 0.18 - 0.21 | [44,45] |
| Musca domestica (housefly) | 3 - 3.6 | ~1000 - 1500 | 1.8 - 2.2 | 35 - 45 | 0.13 - 0.19 | 0.18 - 0.21 | [46] |
| Apis mellifera (honeybee) | 2 - 2.5 | ~1000 - 1200 | 1.57 - 2.09 | 30 - 50 | 0.19 - 0.32 | 0.29 - 0.32 | [47] |
| Eristalis tenax (dronefly) | 3.7 - 3.8 | 1100 - 1400 | 1.87 - 1.90 | 34 | 0.14 - 0.15 | 0.16 | [48] |
| Episyrphus baltealus (hoverfly) | 4.2 - 4.3 | 400 - 540 | 1.15 - 1.47 | 50 | 0.16 - 0.21 | 0.24 - 0.32 | [49] |
| Bombus spp. (bumblebee) | 2.4 - 2.6 | ~1000 | 1.57 | 25 - 35 | 0.24 - 0.27 | 0.22 - 0.28 | [50] |

The frequency scaling described in the previous sections linked natural vortex shedding to active separation control, and then active separation control to insect flapping, via Strouhal number scaling. While separation control and wing flapping are seemingly entirely different phenomena, their comparable Strouhal number scaling can be intuitively understood by the number of vortices shed during a cycle: separation control produces between one and two vortices over the wing at any phase in the forcing cycle, while flapping produces two vortices during the flapping cycle—one on the downstroke and one on the upstroke. The flapping-wing vortices, therefore, are advected more slowly relative to the wing. While the flapping motion generates and advects the LEVs, an axial flow component is required to maintain their stability and prevent bursting [7]. This is produced by the spanwise pressure gradient, that results from the increased wing speed along the span, that induces axial flow to produce stable, conically shaped, helical LEVs.

# 5   Conclusions

This research studied scaling laws for active separation control of constant chord-length, flat-plate wings, based on the generation of maximum lift coefficient. Wind tunnel data were acquired for semispan wings with aspect ratios of 0.75, 1.00, and 1.27, where large-amplitude leading-edge perturbations were introduced using pulsed DBD plasma actuators. Perturbation resulted in a train of leading-edge vortices that substantially enhanced lift. The wind tunnel data were augmented by previously acquired airfoil data, with and without 10% and 20% chord-length Gurney flaps.

The data collapsed best when the dimensionless frequency length-scales were based on the projected streamwise-directed height of the chord length. This scaling was termed the forcing Strouhal number, due to its similarity with the conventional Strouhal number definition based on natural vortex shedding. The link between drag on bluff bodies and lift on flat plate wings was made by recognizing that drag and lift on the latter are linearly proportional. The optimum forcing Strouhal number—namely, that which produced the greatest increases in lift coefficient—was found to be in the range 0.26±0.04 for wings with semispan aspect ratios of 0.75 to ∞ (i.e., airfoils), including airfoils equipped with large Gurney flaps. Flowfield measurements, performed using two-dimensional PIV at the center-span of a finite wing, yielded values for the bubble height and local velocity at separation. This scaling showed that the universal Strouhal number—developed for bluff-body and separation bubble vortex shedding—can be further generalized to active separation control, providing that the natural shedding frequency is substituted by the forcing frequency. The forcing Strouhal number may apply to general wing separation bubbles, but this mainly affects the bubble scaling and does not correlate with either lift or drag.

Insect wing flapping frequencies in hover were examined on the basis of Strouhal number scaling and found to correspond reasonably well to the optimum forcing Strouhal number range. However, accurate angle of attack measurements are required in order to obtain more reliable estimates. Furthermore, to determine if insect flapping scales with the universal Strouhal number, measurements of the local velocities at separation and the height of the LEVs are required. While vortex shedding may be responsible for the initial generation of LEVs, spanwise pressure gradients induce axial flow within the vortices to stabilize them.

# Appendix

Table A1. Summary of insect wing geometry and kinematics for $Re > 400$, copied from the supplemental material of [37].

| Ref. | Species | $R$ (cm) | $\bar{c}$ (cm) | $f_f$ (Hz) | $\Phi$ (rad) | $Re$ | $k_{hf}$ |
|---|---|---|---|---|---|---|---|
| [42] | **Diptera** | | | | | | |
| [42] | Eristalis tenax | 1.12 | 0.298 | 164 | 1.869 | 1,364 | 0.22 |
| [52] | Eristalis tenax | 1.118 | 0.294 | 289 | 1.911 | 2,420 | 0.22 |
| [34] | Eristalis tenax | 1.15 | 0.315 | 183 | 1.501 | 1,327 | 0.29 |
| [34] | Eristalis tenax | 1.15 | 0.315 | 172 | 1.745 | 1,450 | 0.25 |
| [34] | Eristalis tenax | 1.15 | 0.315 | 169 | 1.676 | 1,368 | 0.26 |
| [34] | Eristalis tenax | 1.08 | 0.315 | 172 | 1.71 | 1,334 | 0.27 |
| | **Leidoptera** | | | | | | |
| [53] | Sphinx ligustri | 5 | 1.806 | 30 | 2.09 | 7,549 | 0.27 |
| [35] | Pieris napi | 2.4 | 1.8 | 6 | 2.62 | 905 | 0.45 |
| [35] | Amathes bicolorago | 1.6 | 0.864 | 50 | 2.09 | 1,926 | 0.41 |
| [35] | Macroglossum stellatarum | 2.1 | 0.785 | 73 | 2.09 | 3,353 | 0.28 |
| [35] | Manduca sexta | 5.4 | 1.963 | 27.3 | 2.09 | 8,064 | 0.27 |
| [35] | Manduca sexta | 5 | 1.806 | 29.1 | 2.09 | 7,323 | 0.27 |
| | **Syrphinae** | | | | | | |
| [53] | Syrphus ribesii | 0.82 | 0.243 | 167 | 1.309 | 581 | 0.36 |
| [35] | Syrphus ribesii | 0.9 | 0.267 | 180 | 1.309 | 755 | 0.36 |
| [35] | Syrphus ribesii | 1 | 0.298 | 210 | 1.309 | 1,092 | 0.36 |
| [35] | Syrphus nitens | 0.8 | 0.275 | 172 | 1.309 | 660 | 0.41 |
| | **Hymenoptera** | | | | | | |
| [54] | Apis mellica | 0.98 | 0.291 | 197 | 2.286 | 1,712 | 0.20 |
| [53] | Bombus lapidarius | 1.66 | 0.55 | 143 | 2.6 | 4,526 | 0.20 |
| [55] | Bombus hortorum | 1.41 | 0.419 | 152 | 2.094 | 2,507 | 0.22 |
| [53] | Vespa vulgaris | 1.32 | 0.377 | 143 | 2.09 | 1,983 | 0.21 |
| [35] | Vespa crabro | 2.43 | 0.723 | 104 | 2.09 | 5,092 | 0.22 |